\documentstyle[prl,aps,preprint]{revtex}
\begin{document}
\draft

\newtheorem{defn}{Definition}
\newtheorem{defns}[defn]{Definitions}
\newtheorem{rem}{Remark}
\newtheorem{rems}[rem]{Remarks}
\newtheorem{lem}{Lemma}
\newtheorem{cor}{Corollary}
\newtheorem{prop}{Proposition}
\newtheorem{theo}{Theorem}

\newcommand{\Rn}{{\rm I\!R}} 
\newcommand{\Nn}{{\rm I\!N}} 
\newcommand{\Cn}{{\setbox0=\hbox{
$\displaystyle\rm C$}\hbox{\hbox
to0pt{\kern0.6\wd0\vrule height0.9\ht0\hss}\box0}}} 

\newcommand{\Tr}{{{\bf Tr}}}
\newcommand{\Flim}{{{\lim _{\F_0}\, }}}
\newcommand{\half}{{{1\over 2}}}

\newcommand{\ot}{{\otimes}}
\newcommand{\Hl}{{\cal H}}
\newcommand{\Bi}{{{\cal B}({\cal H}_i )}}
\newcommand{\B}{{{\cal B}({\cal H})}}
\newcommand{\Pc}{{{\cal P}}}
\newcommand{\K}{{\cal K}}
\newcommand{\pn}{{{1 \over {\sqrt n}}}}
\newcommand{\p}{{{\pi_{\varphi}}}}
\newcommand{\s}{{{\varrho^{{1\over 2}}}}}
\newcommand{\si}{{{\varrho^{{1\over 2}}_i}}}
\newcommand{\sig}{{{\sigma^{{1\over 2}}}}}
\newcommand{\St}{{\cal S}}
\newcommand{\M}{{\cal M}}



\title{ Separable and entangled states \\
of composite quantum systems; Rigorous description. }

\author{  Adam W. Majewski}

\address{ Institute of Theoretical Physics and Astrophysics\\
University of Gda\'nsk, 80--952 Gda\'nsk, Poland\\ 
e-mail: fizwam@univ.gda.pl}

\maketitle

\begin{abstract}
 We present a general description of separable states
in Quantum Mechanics.
 In particular, our result gives an easy proof
 that inseparabitity (or entanglement) is a pure quantum
 (noncommutative) notion. This implies that distinction
between separability and inseparabitity
has sense only for composite systems consisting
of pure quantum subsystems. Moreover, we provide the unified
characterization of pure-state entanglement
and mixed-state entanglement.
\end{abstract}

\newtheorem{theorem}{Theorem}
\newtheorem{lemma}{Lemma}
\newtheorem{conclusion}{Corrolary}


Among the most emblematic concepts in Quantum Mechanics
there is
the idea of entanglement. 
Let us remind that this concept enters in description
of quantum correlations between subsystems, so in particular
plays a crucial role in quantum information theory
and quantum computation. This explains why the interest in
better understanding of this concept is so important.

From a formal point of view, a state of a composite quantum system
is called {\it inseparable (or entangled)} if it cannot be represented 
as tensor products of states of its subsystems.
On the contrary, a density matrix describes a {\it separable} state if
it can be expressed as a convex combination
of tensor products of its subsystem states.
These definitions stem from the mathematical
fact that, in general, the convex hull of product states is not a dense
subset
in the state space of the tensor product 
of two von Neumann algebras (cf. \cite{KR}, \cite{Str}).

The purpose of this paper is to discuss,
within the framework of quantum mechanics, 
a general, rigorous description of separable states.
To this end we will need some preliminaries.
Let us consider two physical systems $\Sigma_1$ and $\Sigma_2$
respectively.
As we wish to study quantum mechanical problems,
the systems $\Sigma_i$, $i=1,2$ will be described by
$({\Hl}_i, \Bi, \varrho_i)$, $i=1,2$, where $\Hl_i$ denotes the
set of all pure states of $\Sigma_i$, $\Bi$ the set
of all linear bounded operators on $\Hl_i$ (thus describing
the set of all bounded observables of $\Sigma_i$), and 
finally $\varrho_i$ is a density matrix describing a state of 
$\Sigma_i$. Here and subsequently we shall identify a density
matrix with the notion of state.
In the sequel, we assume that
$\varrho_i$ is an invertible operator (for example, an
equilibrium state of the system $\Sigma_i$).

Let us emphasize that the above assumptions
mean that we restrict ourselves to Dirac's approach
to Quantum Mechanics. 
A more complete theory may be obtained by
an application of a general local quantum physics
approach (cf. \cite{Ha}).
The advantage of using such a general approach
lies in the fact that only within this scheme
one can discuss properly a relation between 
inseparabitity and ``non-locality''.
However, this topic demands more advanced
mathematics and therefore exceeds the scope
of this paper. We wish only to mention that
the presented description 
of separable states can be generalized, 
without any problem, to this general framework.
Furthermore, we shall treat only two-particle entanglement
(cf. \cite{LP}) although a generalization
to multiparcicle entanglement is straightforward too.
 
Let us consider the following family of states
of the complex system $\Sigma = \Sigma_1 + \Sigma_2$
\begin{equation}
\label{1}
 \sigma = \sum_l^{N < \infty} a_l \sigma^{(1)}_l
\ot \sigma^{(2)}_l
\end{equation}

where $a_l \ge 0$, $\sum_l a_l = 1$, and $\sigma_i^{(1)}$
($\sigma_l^{(2)}$) is an arbitrary state of $\Sigma_1$
($\Sigma_2$ respectively).
This family (or more precisely, its closure)
constitutes the set of separable states and it will
be denoted by $\St_{sep}$.
In many recent papers (e.g \cite{We}, \cite{Ho},
\cite{BVSW}, \cite{Pe}) the question of a characterization
of separable states was posed and treated. 
From the existing rigorous results we would like also to mention
the work (\cite{A1}) in which the fine structure of the states
of a composite system was studied in the Hilbert space language.
Here, we want to give
a simple description of $\St_{sep}$ (which
follows from Tomita-Takesaki theory) and
to discuss some physical properties of non-separable states.
For a comprehensive account of Tomita-Takesaki
theory addressed to physicists we refer Haag's book \cite{Ha}
while the mathematical description can be found in \cite{Ar},
and \cite{Ta}.
\vskip 0.5cm
Let $\K_i$ be the set of all Hilbert-Schmidt operators
on the Hilbert space $\Hl_i$ associated with the system $\Sigma_i$.
It is a Hilbert space with the scalar product $(\cdot, \cdot)$ 
of the form $(\sigma, \mu) = Tr \sigma^* \mu$.
Clearly, $\si \in \K_i$.
Now let us form
\begin{equation}
\label{2}
 \Pc_i = closure \{ \varrho^{{1\over 4}} A \varrho^{{1\over 4}}
\} \subseteq \K_i 
\end{equation}

where $A \in \Bi\; \; A \ge0$.
We observe that $\Pc_i$ is the natural cone from Tomita-Takesaki
theory. We note that any operator $ A \in \Bi$
can be considered as a (left)multiplication of $A$
with a Hilbert-Schmidt operator $\mu \in \K_i$.
In such a case the set $\{ A \in \Bi \}$ 
will be denoted by
$\M_i$.  
So we arrived to
\begin{equation}
\label{3}
( \K_i, \M_i, \Pc_i, \si) 
\end{equation}

i.e. we got the standard form of $\Bi$ in 
the framework of Tomita-Takesaki theory.
The reader non acquainted with this theory can consider
$( \K_i, \M_i, \Pc_i, \si)$ as the following quadruple:
(all Hilbert-Schmidt operators on $\Hl_i$, $\Bi$ acting on the set
of Hilbert-Schmidt operators as left multiplications,
all positive Hilbert-Schmidt operators, square-root
of the distinguished density matrix) where $\Hl_i$ is the Hilbert space
of all pure states of the system $\Sigma_i$. 
We wish to recall one important result of the just mentioned theory.
Namely, each state $\sigma_i$ of $\Sigma_i$ 
can be represented (in the unique way) by a vector 
$\sig_i \in \Pc_i$ such that
\begin{equation}
\label{4}
<A_i>_{\sigma_i} \equiv Tr A_i \sigma_i = ( \sig_i, A_i \sig_i)
\end{equation}

where $\sigma_i$ is a state of the system $\Sigma_i$,
and with the small abuse of notation
we used the same letter $A_i$ for 
the operator  in
$\Bi$ (left hand side of (4)) 
representing an arbitrary but fixed observable
and for the operator in $\M_i$ (right hand side of (4))
which represents the same observable.
\vskip 0.5cm
Now let us turn to the promised characterization
of separable states of the composite system $\Sigma$.
Using the prescription that the Hilbert space $\Hl$ of all pure
states of a composite system $\Sigma$
is described by the tensor product of Hilbert spaces
$\Hl_i$ of pure states of its components, $\Hl = \Hl_1 \ot \Hl_2$,
we infer that the standard form associated with $\Sigma$
is 
\begin{equation}
\label{5}
(\K_1 \ot \K_2, \M_1 \ot \M_2, \Pc, \s_1 \ot \s_2 ) 
\end{equation}

where $\Pc = closure \{ \varrho^{{1\over 4}}_1 \ot \varrho^{{1\over 4}}_2
A \varrho^{{1\over 4}}_1 \ot \varrho^{{1\over 4}}_2,
\;\; A \in \M_1 \ot \M_2, \;\; A\ge 0 \} \subset \K_1 \ot \K_2$.
It is important to note that $\Pc_1 \ot \Pc_2 \equiv closure
\{ \sum_k a_k x^{(1)}_k \ot x^{(2)}_k, \;\; a_k \ge 0, \sum_k a_k = 1,
\;\; x^{(i)}_k \in \Pc_i \} $ is, in general, a proper subset of $\Pc$.
This is a result of the fact that the set of all positive 
operators in $\M_1 \ot \M_2$ contains, in general as a proper subset, 
the closure of $\sum A^{(1)}_k \ot A^{(2)}_k$ where 
$A^{(1)}_k$ is a positive operator in $\M_1$ while
$A^{(2)}_k$ is a positive operator in $\M_2$.
(For a discussion of this fact in physical terms see
\cite{We} while for mathematical argument see \cite{Cu}).

Consider the tensor product $\M_1 \ot \M_2$. 
Our characterization of separable states starts with
the observation that each convex combination
$\omega_0 = \sum_i \lambda_i \omega^{(1)}_i \ot \omega^{(2)}_i$
of product states of $\M_1 \ot \M_2$ can be expressed in the form
(cf. (4))
\begin{equation}
\label{6}
\omega_0 = \sum_i \lambda_i \omega_{x_i} \ot \omega_{y_i}
\end{equation}
where $\omega_{x_i}(\cdot) = (x_i, \cdot x_i)$,
$x_i \in \Pc_1$, 
 $\omega_{y_i}(\cdot) = (y_i, \cdot y_i)$,
$y_i \in \Pc_2$. (We recall that $(\cdot, \cdot)$
is the scalar product in $\cal K$).
Thus for $A \in \M_1$, $B \in \M_2$
\begin{equation}
\label{7}
\omega_0(A \ot B) = \sum_i \lambda_i (x_i,A x_i)(y_i,By_i)
= \sum_i \lambda_i (x_i \ot y_i, (A \ot B) \cdot x_i \ot y_i)
\end{equation}
Hence
\begin{equation}
\label{8}
\omega_0(A\ot B) = Tr \varrho_0 A \ot B
\end{equation}
where 
\begin{equation}
\label{9}
\varrho_0 = \sum_i \lambda_i \vert x_i \ot y_i><x_i \ot y_i \vert 
\end{equation}
In (9) we have written projectors in the Dirac's notation, i.e.
$\vert x_i \ot y_i><x_i \ot y_i \vert  \vert f \ot g>
\buildrel \rm def \over = <x_i \ot y_i \vert  f \ot g>
\vert x_i \ot y_i> \equiv (x_i \ot y_i,  f \ot g)_{\K}
\vert x_i \ot y_i> $ 
for any $f\in \K_1$ and $g \in \K_2$. 
Consequently, we have the following characterization
of separable states:

 {\it The set of separable states is the norm
closure of the set
\begin{equation}
\label{10}
\{ \varrho_0 = \sum_i \lambda_i
\vert x_i \ot y_i><x_i \ot y_i \vert \} 
\end{equation}
where $x_i \in \Pc_1$, $y_i \in P_2$, $i=1,2,...$}

\vskip 0.5cm
To avoid the future confusion we emphasize
that the set of vectors $\{ \vert x_i \ot y_i > \}$
does not form, in general, an orthogonal system in $\K_1 \ot \K_2$.
Consequently, $(10)$ is not a spectral resolution of 
the corresponding density matrix.
\skip 0.5cm
Now let us rewrite (7) in the following way
\begin{equation}
\label{11}
 (\s_0, (A \ot B) \s_0) \equiv
\omega_0 (A \ot B) = \sum_i \lambda_i
(x_i \ot y_i, (A \ot B) x_i \ot y_i) 
\end{equation}
where $\s_0 \in \Pc$ while $A$ ($B$) is an arbitrary
operator in $\M_1$ ($\M_2$, respectively).
We remind the reader that (9) is not a spectral decomposition
of $\varrho_0$. Hence, $\s_0$, being the representative vector of
$\omega_0$ in $\Pc$, is not the square root of $\varrho_0$
(given by (9)) in the sense 
that it is not equal to
$\sum_i \lambda_i^{1\over 2}
\vert x_i \ot y_i><x_i \ot y_i \vert$. 
Equality (11) implies
\begin{equation}
\label{12}
 Tr P_{\s_0} A \ot B = \sum_i
\lambda_i Tr P_{x_i \ot y_i} A \ot B
\end{equation}
Here, $ P_{\s_0}$ and $P_{x_i \ot y_i}$
are projectors in ${\cal B}(\K)$. Clearly,
$Tr$ is taken with respect to a basis in $\K$.
So
\begin{equation}
\label{13}
P_{\s_0} = \sum_i \lambda_i P_{x_i \ot y_i}.
\end{equation}
Hence
\begin{equation}
\label{14}
\s_0 = \sum_i \lambda_i P_{x_i \ot y_i} \s_0
= \sum_i \lambda_i (x_i \ot y_i, \s_0) x_i \ot y_i
\end{equation}
We note that
$$(x_i \ot y_i, \s_0) \ge 0 $$
as $\Pc_1 \ot \Pc_2 \subset \Pc $ and $ \Pc$ is a selfdual cone.
Consequently
\begin{equation}
\label{15}
\s_0 \in \Pc_1 \ot \Pc_2.
\end{equation}
Conversely, let $0 \ne\s \in \Pc_1 \ot \Pc_2$, i.e.
\begin{equation}
\label{16}
\s = \sum_i \lambda^0_i x_i \ot y_i
\end{equation}
with $0\ne x_i \in \Pc_1$, $0\ne y_i \in \Pc_2$
and $\lambda^0_i >0$. This is an easy direction,
however for the reader convenience we present
all necessary details. We observe
\begin{equation}
\label{17}
(\s, x_i \ot y_i) >0
\end{equation}
for all $i$. To prove this inequality let us assume
\begin{equation}
\label{18}
(\s, x_j \ot y_j)=0 
\end{equation}
for some fixed $j$. Thus
\begin{equation}
\label{19}
0 = \sum_k \lambda_k^0 (x_k \ot y_k, x_j \ot y_j)
\end{equation}
where $(x_k \ot y_k, x_j \ot y_j) \ge 0$ by selfduality of the natural cone.
So each term of the sum has to be equal to $0$.
In particular
\begin{equation}
(x_j \ot y_j, x_j \ot y_j) = \Vert x_j \ot y_j \Vert^2 =0
\end{equation}
So $x_j \ot y_j = 0$ and this is a contradiction. Therefore
\begin{equation}
 (\s, x_i \ot y_i) >0
\end{equation}
for each $i$. Consequently, we can rewrite (16) in the following way
\begin{equation}
\label{20}
\s = \sum_i \lambda^0_i x_i \ot y_i 
= \sum_i {{\lambda^0_i} \over (x_i \ot y_i, \s)}
(x_i \ot y_i, \s) x_i \ot y_i
\equiv \sum_i \lambda_i (x_i \ot y_i, \s) x_i \ot y_i
\end{equation}
where $ \lambda_i = { \lambda^0_i \over
(x_i \ot y_i, \s)} >0$.
In other words, we arrived to the same form for $\s$ as in (14).
This justifies the following equality
\begin{equation}
\label{21}
P_{\s} = \sum_i \lambda_i P_{x_i \ot y_i}
\end{equation}
and therefore $P_{\s}$ determines the separable state.
More precisely, {\it we use the one-to-one correspondence
between a product state $\sigma \ot \mu$ and a simple tensor
$x_{\sigma} \ot y_{\mu} \in \Pc$}. Subsequently, 
this correspondence is 
extended by convexity.
To summarize, {\it we have obtained the one-to-one
correspondence between the set of normalized vectors in 
$\Pc_1 \ot \Pc_2$ and the set of 
all separable states $\St_{sep}$}. 
It should be noted however that the above 
correspondence is implemented by the proper choice of projectors (cf. (13)).
\vskip 1cm
\leftline{{\it Example:}}
To illustrate the presented description let us consider
the following example.
Let $\eta = \lambda_1 x \ot y + \lambda_2 y \ot x$
be a vector in $\Hl \equiv \Hl_1 \ot \Hl_2$, $x \in \Hl_1$,
$y \in \Hl_2$, $\lambda_1, \lambda_2 \in \Cn$.
We note that the vector $\eta$ is a prototype
of an entangled (pure) state in Dirac's quantum mechanics.
Obviously, this state is uniquely determined by the projector
$P_{ \lambda_1 x \ot y + \lambda_2 y \ot x}$ which
belongs to $\Pc$. However, easy but a little bit tedious
calculations (which we left to the reader) show that
$P_{ \lambda_1 x \ot y + \lambda_2 y \ot x} \not\in \Pc_1 \ot \Pc_2$.
Consequently, we have shown that the standard example
of pure entangled state fits well in our characterization.

To see peculiarity of the subset $\Pc_1 \ot \Pc_2$ of $\Pc$
(which {\it uniquely} characterizes the set of all separable states)
let us write down explicitly
an operator $\sigma$ such that
\begin{equation}
\label{22}
(\sigma, P_x \ot P_y) \ge 0
\end{equation}
for any $x \in \Hl_1$, $y \in \Hl_2$ while $\sigma \not\in \Pc$. 
This clearly shows that the set $\Pc_1 \ot \Pc_2$
does not have the selfduality property. We remind the reader
that this is extremely important property of the natural
cone $\Pc$ (which describes the set of {\it all} states).
Let us consider (cf. \cite{Cu})
\begin{eqnarray}
\label{23}
\sigma = (\vert f><f \vert) \ot (\vert g><g \vert)
+(\vert g><g \vert) \ot (\vert f><f \vert)\\
- (\vert f><g \vert) \ot (\vert f><g \vert)
- (\vert g><f \vert) \ot (\vert g><f \vert)\nonumber
\end{eqnarray}
where $f,g \in \Hl_1 = \Hl_2 \equiv \Hl_0$ and $(f,g) =0$.
Here, $\vert g><f \vert \equiv (g,h)_{\Hl_0} \vert f>$. 
One can show that
\begin{equation}
\label{24}
(\sigma, \xi)_{\K} \ge 0
\end{equation}
for every $\xi \in \Pc_1 \ot \Pc_2$ while there does exit
an element $\theta$ in $\Pc$
\begin{eqnarray}
\label{25}
\theta = (\vert f><f \vert) \ot (\vert f><f \vert)
+ (\vert g><f \vert) \ot (\vert g><f \vert)\\
+(\vert f><g \vert) \ot (\vert f><g \vert)
+(\vert g><g \vert) \ot (\vert g><g \vert)\nonumber
\end{eqnarray}
such that 
\begin{equation}
\label{26}
(\sigma, \theta)_{\K} <0
\end{equation}
Thus the set $\Pc_1 \ot P_2$ is a (proper) subset of $\Pc$ 
(provided that $dim \Hl_1 \ge 2$, $dim \Hl_2 \ge2$) 
and therefore  $\Pc_1 \ot P_2$
is too a small set to describe all states of composite quantum system.
\smallskip
\rightline{$\circ$}
\smallskip
\vskip 1cm

The principal significance of this characterization
is that it offers a clear 
explanation why the set of separable states of a general 
complex system is a proper subset 
of the set of all states. The characterization 
of the set of separable states ${\cal S}_{sep}$ needs only a 
proper subset $\Pc_1 \ot \Pc_2$ of the natural cone $\Pc$ while for the 
description of all states of $\Sigma$ we have to use the whole
set $\Pc$. In other words, {\it convex combinations of product states
do not lead to the set of all states}.
To get another important and somewhat surprising 
consequence of this characterization we recall the following
result (see \cite{Cu}) saying that $\Pc = \Pc_1 \ot \Pc_2$
if and only if  $\Sigma_1$ or $\Sigma_2$ (or both)
is a classical subsystem, i.e. the algebra of observables of
$\Sigma_i$ is an abelian algebra.
This means that for such a system {\it the set of separable
states is equal to the set of all states}.
In particular, this implies that any approximation
of a real quantum atom interacting with the quantum electromagnetic
field by a system of quantum atom and classical electromagnetic
field {\it kills the distinction between
separability and inseparabitity}.
For an interpretation of this result, in terms
of correlations, see below. 
As another example in this direction let us mention
a system in quantum information theory where one
of channels is assumed to be a classical one. 
Again, for such a model the set of all separable states is
equal to the set of all states.
Finally we want to note that our characterization
partly clarifies the theory of mixed-state
entanglement (cf. \cite{BVSW}). Namely, we present
the unified characterization
of ``pure-state entanglement'' and mixed-state entanglement''.
For other consequences of the presented characterization see
\cite{KM}.

\vskip 0.7cm
Now let us turn to the following question: What is the basic 
``physical'' difference
between the sets of separable and non-separable
states? To answer this question let us recall another important
property of a natural cone $\Pc$. Let $\xi$ be a vector 
in a fixed natural cone $\Pc$ such that $\{ A \xi, \; A \in
\M_1 \ot \M_2 \}$ is dense in $\K_1 \ot \K_2$.
Then a natural cone constructed with such
a vector $\xi$ is equal to the previous one.
Let us discuss this property in physical terms.
Assume $\varrho_i$ is an equilibrium state for $\Sigma_i$, so
$\varrho_i = Z_i^{-1} exp\{ - \beta H_i\}$
where $Z_i$ is the partition function, $H_i$ the hamiltonian
of the system $\Sigma_i$, and $\beta$ stands for 
the inverse temperature.
Now it is clear that the set $\Pc_1 \ot \Pc_2$ is built on 
uncorrelated state $\varrho_1 \ot \varrho_2$ in such a way that
it does not 
contain essential quantum correlations
 (we remind the reader that to form $\Pc_1 \ot \Pc_2$ we
need only to consider convex combinations of $x_i \ot y_i$,
$x_i \in \Pc_1$, $y_2 \in \Pc_2$). To form $\Pc$ we can use
a Gibbs state of the form $Z^{-1} exp \{ -\beta (H_1 + H_2 + H_I) \} $
where $H_I$ describes any interaction between
$\Sigma_1$ and $\Sigma_2$. So $\Pc$ contains,
from the very beginning, also
vectors describing all quantum correlations.
Clearly, this feature can be also explained by the fact
that the set $\{ A \in \M_1 \ot \M_2, \; A \ge 0 \}$
contains all (positive) ``correlated'' observables.
Consequently we arrive to the following conclusion:
{\it the set of nonseparable states 
contains pure quantum correlations of the 
complex system $\Sigma$, while separable states do not have
this property}.
\vskip 0.5cm
We wish close this note with a remark that
the just given characterization of states 
can be used in the debate on the Einstein-Podolsky-Rosen
paradox and Bell's inequality (for a deeper discussion on these subjects
we refer the reader to (\cite{A2}). 
Namely,
considering a composite interacting system $\Sigma = \Sigma_1 + \Sigma_2$
in a nonseparable state $\sigma$ it is difficult to
specify the notion of subsystem.
The main difficulty in carrying out such
a specification is that such
a procedure should take into account 
"causal ties" which are hidden in $\sigma$ (cf. \cite{Ha}). 
Of course, this problem does not exist for
the pair $(\Sigma, \rho)$ where $\rho \in \St_{sep}$.

\vskip 1 true cm

{\bf Acknowledgments}: The author would like to thank 
Luigi Accardi and Nobuaki Obata
for hospitality at Graduate School of Polymathematics of Nagoya University.

The author would like also to acknowledge
the partial supports of BW/5400-5-0303-7 and the Nagoya University.


\end{document}